\pdfoutput=1
\documentclass[aps,prl,preprint,superscriptaddress,tightenlines,nofootinbib]{revtex4}
\usepackage{graphicx}
\usepackage{amsmath,amssymb,latexsym}
\usepackage{bm}
\usepackage{epsfig}
\newcommand{\PRE}[1]{{#1}}   

\newcommand{\postscript}[2]{\setlength{\epsfxsize}{#2\hsize}
   \centerline{\epsfbox{#1}}}

\begin{document}
\title{Neutrino Cosmology after WMAP and LHC7}

\author{Luis Alfredo Anchordoqui}
\affiliation{Department of Physics,\\ University of Wisconsin-Milwaukee,
P.O. Box 413, Milwaukee, WI 53201, USA
\PRE{\vspace*{.1in}}
}

\author{Haim Goldberg}
\affiliation{Department of Physics,\\
Northeastern University, Boston, MA 02115, USA
\PRE{\vspace*{.1in}}
}

\date{November 2011}
\PRE{\vspace*{.5in}}
\begin{abstract}\vskip 3mm\noindent
  The gauge-extended $U(1)_C \times SU(2)_L \times U(1)_{I_R} \times
  U(1)_L$ model has the attractive property of elevating the two major
  global symmetries of the standard model (baryon number $B$ and
  lepton number $L$) to local gauge symmetries.  The $U(1)_L$ symmetry
  prevents the generation of Majorana masses, leading to three
  superweakly interacting right-handed neutrinos. This also renders a
  $B-L$ symmetry non-anomalous. We show that the
  superweak interactions of these Dirac states (through their coupling
  to the TeV-scale $B-L$ gauge boson) permit right-handed neutrino
  decoupling just above the QCD phase transition: $175~{\rm MeV}
  \lesssim T_{\nu_R}^{\rm dec} \lesssim 250~{\rm MeV}$.  In this
  transitional region, the residual temperature ratio
  between $\nu_L$ and $\nu_R$ generates extra
  relativistic degrees of freedom at BBN and at the CMB epochs. 
  Consistency (within $1\sigma$) with
  both WMAP 7-year data and the most recent estimate of the primordial
  $^4$He mass fraction is achieved for $3~{\rm TeV} <
  M_{B-L} < 6~{\rm TeV}$. The model is fully predictive, and can be
  confronted with dijet and dilepton data (or lack thereof) from LHC7
  and, eventually, LHC14.
\end{abstract}

\maketitle

Heavy neutral vector gauge bosons ($Z'$'s) are ubiquitous in extensions of
the standard model (SM)~\cite{Langacker:2008yv},  often including a gauged $B-L$ symmetry.
This $U(1)$ symmetry is non-anomalous if the
three left-handed Weyl neutrinos are accompanied by three right-handed
neutrinos. The $Z'$ masses are {\em a priori} open parameters -- not
determined by the low energy effective theory -- but subject to recent experimental 
bounds ($M_{Z'} \agt 3~{\rm TeV}$) from searches of
dilepton~\cite{Chatrchyan:2011wq} and dijet~\cite{Chatrchyan:2011ns}
events in the 7~TeV run of the Large Hadron Collider (LHC7). In this
Letter we re-examine some critical cosmological issues surrounding
the presence of the six additional neutrino degrees of
freedom~\cite{Steigman:1979xp} correlated to the presence of a $Z'$ 
in our dynamical model which is coupled to $B-L$. These considerations, when viewed in
the context of most recent data collected by the Wilkinson Microwave
Anisotropy Probe (WMAP)~\cite{Komatsu:2010fb}, are found to constrain the mass
of the $Z'$ to an interestingly narrow band, which will be directly probed by LHC14.

For a good part of the past two decades, big-bang nucleosynthesis
(BBN) provided the best inference of the radiation content of the
universe. The time-dependent quantity being the neutron abundance at
$t \agt \tau_n$, which regulates the primordial fraction of baryonic
mass in $^{4}$He,
\begin{equation}
Y_{\rm p} \simeq 0.251 + 0.014 \ \delta N_\nu^{\rm eff} + 0.0002 \ \delta \tau_n + 
0.009 \ \ln \left(\frac{\eta}{5 \times 10^{-10}} \right)\,,
\end{equation}
where $\delta N_\nu^{\rm eff} = N_\nu^{\rm eff} -3$ is the effective
number of extra (non-SM) light neutrino species, $\tau_n$ is the
neutron half life, and $\eta$ is the ratio of the baryon number
density to the photon number density~\cite{Sarkar:1995dd}. The
observationally-inferred primordial fractions of baryonic mass in
$^{4}$He ($Y_{\rm p} = 0.2472 \pm 0.0012$, $Y_{\rm p} = 0.2516 \pm
0.0011$, $Y_{\rm p} = 0.2477 \pm 0.0029$, and $Y_{\rm p} = 0.240 \pm
0.006$~\cite{Izotov:2007ed}) have been constantly favoring $N_\nu^{\rm
  eff} \alt 3$~\cite{Simha:2008zj}.  Out of the blue, two recent
independent studies determined $Y_{\rm p} = 0.2565 \pm 0.001 ({\rm
  stat}) \pm 0.005 ({\rm syst})$ and $Y_{\rm p} = 0.2561 \pm
0.011$~\cite{Izotov:2010ca}. For $\tau_n = 885.4 \pm 0.9~{\rm s}$ and
$\tau_n = 878.5 \pm 0.8~{\rm s}$, the updated effective number of
light neutrino species is reported as $N_\nu^{\rm eff} =
3.68^{+0.80}_{-0.70}$ ($2\sigma$) and $N_\nu^{\rm eff} =
3.80^{+0.80}_{-0.70}$ ($2\sigma$), respectively.

Very recently, in support of these trends, observations of the cosmic microwave background (CMB)
anisotropies and the large-scale structure distribution have allowed
a probe of $N_\nu^{\rm eff}$ at the CMB decoupling epoch with
unprecedented precision. The relativistic particles that stream freely
influence the CMB in two ways: {\it (1)} their energy density alters the
matter-radiation equality epoch, and {\it (2)} their anisotropic
stress acts as an additional source for the gravitational potential
via Einstein's equations. Hence, the number of light relativistic
species becomes a function of the matter density $(\Omega_m
h^2)$ and the redshift of matter-radiation equality $(z_{\rm eq})$,
\begin{equation}
1 + z_{\rm eq} = \frac{\Omega_m h^2}{\Omega_r h^2} =  \frac{\Omega_m h^2}{\Omega_{\rm \gamma} h^2}
 \left[ 1 + \frac{7}{8} \, \left( \frac{4}{11} \right)^{4/3} N_\nu^{\rm eff} \right]^{-1} \,,
\end{equation}
where $\Omega_\gamma h^2 = 2.469 \times 10^{-5}$ is the present-day
photon energy density (for $T_{\rm CMB} = 2.725~{\rm K}$) and the
scaled Hubble parameter $h$ is defined by $H = 100 \, h~{\rm km} \, {\rm
  s}^{-1} \, {\rm Mpc}^{-1}$~\cite{Kolb:1990vq}. The variation in $N_\nu^{\rm eff}$ reads
\begin{equation}
\frac{\delta N_\nu^{\rm eff}}{N_\nu^{\rm eff}} \simeq 2.45 \ \frac{\delta(\Omega_m h^2)}{\Omega_m h^2}
 - 2.45 \ \frac{\delta z_{\rm eq}} { 1 + z_{\rm eq}} \, .
\end{equation}
The equality redshift is one of
the fundamental observables that one can extract from the CMB power
spectrum. More specifically, WMAP data constrain $z_{\rm eq}$ mainly
from the height of the third acoustic peak relative to the first
peak~\cite{Komatsu:2010fb}. The fractional error in $\Omega_m h^2$ is
determined using external data: the latest distance measurements from
the Baryon Acoustic Oscillations (BAO) in the distribution of
galaxies~\cite{Percival:2009xn} and precise measurements of the Hubble
constant $H_0$~\cite{Riess:2009pu}. The parameter constraints from the
combination of WMAP 7-year data, BAO, and $H_0$ lead to 
$N_\nu^{\rm eff} = 4.34^{+0.86}_{-0.88}~(68\% {\rm
  CL})$~\cite{Komatsu:2010fb}.

In summary, though uncertainties remain large, the most recent
cosmological observations show a consistent preference for additional
relativistic degrees of freedom (r.d.o.f.) during BBN and the CMB
epochs. We take these hints as motivation for our analysis, which
consists of the following tasks: {\em (1)} to present a model in which
the additional r.d.o.f. are three flavors of light right-handed
neutrinos which interact with the SM fermions via the exchange of
heavy vector fields; {\em (2)} to suppress the six additional
fermionic r.d.o.f. to levels in compliance with BBN and CMB. This is
accomplished by imposing the decoupling of $\nu_R$'s from the plasma
{\em early enough} so that they undergo incomplete reheating during
the QCD phase transition; and {\em late enough} so as to leave an
excess neutrino density suggested by the
data~\cite{Feng:2011uf}. These requirements strongly constrain the
masses of the heavy vector fields. Together with the couplings, which
are determined in accord with other considerations, the model is fully
predictive, and can be confronted with dijet and dilepton data (or
lack thereof) from LHC7 and, eventually, LHC14.

An economic choice of the  model to implement the task outlined above is based on the gauge-extended sector $U(3)_C \times SU(2)_L \times U(1)_{I_R} \times U(1)_L$~
\cite{Anchordoqui:2011eg}.  The resulting $U(1)$ content gauges the baryon number $B$ [with $U(1)_B \subset U(3)_C $], the lepton number $L$, and a third additional 
abelian charge $I_R$  which acts as the third isospin component of an $SU(2)_R$.  The usual electroweak hypercharge  is a linear combination of these three $U(1)$ 
charges:  $Y = \frac{1}{2} (B- L) + I_{R}$.  The matter fields consist of six sets (labeled by an index $i = 1 - 6$) of Weyl fermion-antifermion pairs: $\left( U_R,\, D_R,\, 
L_L,
\, E_R,\, Q_L,\, N_R\right)$. The field $N_R$ is the right-handed neutrino (and left-handed antineutrino) accompanying the fields in the set $E_R$, with mass $\sim 1~
{\rm 
eV}.$ The gauging of lepton number precludes the presence of a seesaw for generating Majorana neutrino masses.  In addition to the SM interactions, these fields experience two $U(1)$ gauge interactions mediated by two associated vector bosons ($Z'$ and $Z''$) whose 
masses 
lie well above a TeV.

The initially free parameters consist of three couplings $g_B, g_L,g_{I_R}.$ These are augmented by three Euler angles to allow for a field rotation to coupling diagonal 
in 
hypercharge. This diagonalization fixes two of the angles and the orthogonal nature of the rotation introduces one constraint on the couplings $P(g_Y,g_B,g_L,g_{I_R}) 
=0$. The baryon number coupling $g_B$ is fixed to be $\sqrt{3/2}$ of the non-abelian $SU(3)$
coupling at the scale of $U(N)$ unification, and is therefore determined at all energies
through RG running. This leaves one free angle and two couplings with one constraint.
The two remaining degrees of freedom allow a further rotation leaving $Z'$ to couple to $B$ at 90\% and $Z''$ to couple to $B-L$ at 99\%.\footnote{ Although not
  generally appreciated, it is important to note that a 100\% coupling
  of the $Z'$ and $Z''$ to $B$ and $B-L$, respectively, is possible
  only if the $U(1)$ gauge coupling constants are equal.} The $U(1)$
quantum numbers and the physical couplings of the $Z'$ and $Z''$ to
the fermion fields are given in Table~\ref{t:spectrum}.  These
couplings, which have been computed
elsewhere~\cite{Anchordoqui:2011eg}, are functions of the charge
assignments, the $U(1)$ gauge couplings, and the mixing angles. All
fields in a given set have a common $g',\, g''$ couplings.\footnote{Our couplings are consistent with the bounds
  presented in~\cite{Williams:2011qb} from a variety of experimental constraints.}

The  model as described enjoys distinct advantages: {\em (1)}~Gauging of the anomalous $B$ and its cancellation by generalized Green-Schwarz mechanism (which 
leaves 
$B$ as a global symmetry) prevents proton decay. {\em (2)}~The presence of $N_R$ renders $B-L$ non-anomalous. This has been appealing for minimal extension of 
SM 
at the TeV-scale.
For example, the mass growth of $Z''$ can occur via a conventional Higgs mechanism at TeV without relying on possible Planck scale physics. {\em (3)}~By 
inspection of 
Table~\ref{t:spectrum} the charges $B$, $L$, and $I_R$ are mutually orthogonal in the fermion space. This will maintain the othogonality relation $P=0$ to one loop without inducing kinetic 
mixing~\cite{Anchordoqui:2011eg}.

We begin by first establishing, in a model independent manner, the range of
decoupling temperatures implied by the the BBN and CMB analyses. For this work, the physics of interest will be taking place at energies in the
region of the QCD phase transition, so that we will restrict ourselves to the following
fermionic fields, and their contribution to r.d.o.f.:
\begin{equation}
\left[ 3 u_R \right]  + \left[ 3 d_R \right] + \left[ 3 \nu_L + e_L + \mu_L \right]
+ \left[ e_R + \mu_R \right]+ \left[ 3 u_L + 3d_L \right]+ \left[ 3
  \nu_R \right] \, .
\end{equation}
This amounts to 22 Weyl fields, translating to 44 fermionic r.d.o.f.

\begin{table}
\caption{Quantum numbers of chiral fermions and their couplings to $Z'$ and $Z''$ gauge bosons.}
\begin{tabular}{c|ccccc|cc}
\hline
\hline
 Name &~~Representation~~& ~~~~~$B$~~~~~ & ~~~~~$L$~~~~~~ & ~~~$I_R$ & ~~~~~~~$Y$~~~~ &~~ $g'$~ &~~$g''$\\
\hline
~~$U_R$~~ & $( 3,1)$ &   $\frac{1}{3}$ & $0$ & $\phantom{-}\frac{1}{2}$ & $\phantom{-}\frac{2}{3}$ & ~~~$0.368$~~~ & ~~~$- 0.028$~~~ \\[1mm]
~~$D_R$~~ &  $( 3,1)$&    $\frac{1}{3}$ & $0$ & $- \frac{1}{2}$ & $-\frac{1}{3}$  &  ~~~$0.368$~~~ &  ~~~$- 0.209$~~~ \\[1mm]
~~$L_L$~~ & $(1,2)$&    $0$ &  $1$ & $\phantom{-}0$ & $-\frac 1 2$  & ~~~$0.143$~~~ & ~~~$\phantom{-} 0.143$~~~\\[1mm]
~~$E_R$~~ &  $(1,1)$&   $0$ & $1$ &  $- \frac{1}{2}$ & $- 1$ & ~~~$0.142$~~~ & ~~~$\phantom{-} 0.262$~~~\\[1mm]
~~$Q_L$~~ & $(3,2)$& $\frac{1}{3}$ & $0 $ & $\phantom{-} 0$ & $\phantom{-}\frac{1}{6}$    &  ~~~$0.368$~~~ & ~~~$-0.119$~~~ \\[1mm]
~~$N_R$~~&  $(1,1)$&    $0$ & $1$ &  $\phantom{-} \frac{1}{2}$ & $\phantom{-} 0$ & ~~~$0.143$~~~ & ~~~$\phantom{-}0.443$~~~\\[1mm]
\hline
\hline
\end{tabular}
\label{t:spectrum}
\end{table}

Next, in line with our stated plan, we use the data estimate to calculate the range of decoupling temperature. The effective number of neutrino species contributing to 
r.d.o.f. can be written as
\begin{equation}
N_\nu^{\rm eff} = 3 \left[1 + \left(\frac{T_{\nu_R}}{T_{\nu_L}}\right) ^4\right] \,;
\end{equation}
therefore, taking into account the isentropic heating of the rest of the plasma between $\nu_R$ decoupling temperature $T_{\rm dec}$ and the end of the reheating phase,
\begin{equation}
\delta N_\nu^{\rm eff} = 3 \left(\frac{N(T_{\rm end})}{N(T_{\rm dec})} \right)^{4/3} \,,
\label{7}
\end{equation}
where  $T_{\rm end}$ is the temperature at the end of the reheating phase, and $N(T) = r(T)(N_{\rm B} + \frac{7}{8} N_{\rm F})$ is the effective number of r.d.o.f.  at 
temperature 
$T$, with $N_{\rm B} = 2$ for each real vector field and $N_{\rm F} = 2$ for each 
spin-$\frac{1}{2}$ Weyl field. The coefficient $r(T)$ is unity for the lepton and photon contributions, and is the ratio $s(T)/s_{\rm SB}$ for the quark-gluon plasma. Here $s(T) (s_{\rm SB})$ is the actual (ideal Stefan-Bolzmann) entropy. Hence  $N(T_{\rm dec})
 = 37~r(T_{\rm dec}) + 14.25.$ We take $N(T_{\rm end}) = 10.75$ reflecting
$(e_L^-+e_R^+ + e_R^- + e_L^+ \nu_{eL} + \bar \nu_{eR} + \nu_{\mu L} + \bar \nu_{\mu R} + \nu_{\tau L} + \bar \nu_{\tau R} + \gamma_L + \gamma_R)$. We 
consistently 
omit $\nu_R$ in considering the thermodynamics part of the discussion, but will include it  when dealing with expansion.
As stated in the introduction
\begin{equation}
\delta N_\nu^{\rm eff} = \left \{ \begin{array}{ccl} 0.68^{+0.40}_{-0.35} & ~~(1 \sigma) & ~~{\rm BBN} \\
1.34^{+0.86}_{-0.88} & ~~(1\sigma) & ~~{\rm WMAP + BAO +} H_0
\end{array} \right.
\label{8}
\end{equation} 
so the excess r.d.o.f.  will lie within $1 \sigma$ of the central value of each experiment if
$0.46 < \delta N_\nu^{\rm eff} < 1.08$. From Eqs.~(\ref{7}) and (\ref{8}), the allowable range for $N$ is 
\begin{equation}
23 < N (T_{{\rm dec}} )< 44 \, .
\end{equation}
This is achieved for $0.24 < r(T_{\rm dec}) < 0.80.$ By comparing to Fig.~8 in Ref.~\cite{Bazavov:2009zn}, this can be translated into temperature range 
\begin{equation}
175~{\rm MeV} < T_{{\rm dec}} < 250~{\rm MeV} \, ,
\label{10}
\end{equation}
with the lower temperature coinciding with the region of most rapid rise of the entropy.  Thus, the data implies that the $\nu_R$ decoupling takes place during QCD phase transition. 

We now turn to use our model in conjunction with the decoupling condition to constrain its parameters. To this end we calculate the interaction rate
$\Gamma(T)$ for a right-handed neutrino and determine $T_{\rm dec}$ from the plasma via the prescription
\begin{equation}
 \Gamma(T_{\rm dec}) = H(T_{\rm dec}) \, .
\label{haim1}
\end{equation}
Let $f_L^i$ be a single species of Weyl fermion, representing the two r.d.o.f. 
$\{f_L^i, \bar f_R^i \}$, where the superscript indicates  bins $i = 3,5$. Similarly 
$f_R^i \in  \{f_R^i, \bar f_L^i\},$ for $i = 1,2,4,6$. Notice that the subscripts $L,R$ 
denote the actual helicities of the massless particles in question, not the chirality of the fields. With this 
said, we may write the amplitude for $f_L^i$ scattering
\begin{equation}
\mathfrak{M} \left(\nu_R(p_1) f_L^i (p_2) \to \nu_R (p_3) f_L^i(p_4)\right) 
= \frac{G_i}{\sqrt{2}} [\bar u(p_3)  \gamma^\mu  (1+\gamma_5)  u(p_1)] [\bar u(p_4)  \gamma_\mu  
(1- \gamma_5)  u(p_2) ] \, .
\end{equation}
The other 3 amplitudes are obtained by the crossing substitutions in the second square bracket; for scattering from
\begin{eqnarray}
\bar f_R^i & \to & \bar v(p_2) \, \gamma_\mu \, (1- \gamma_5) \, v(p_4) \nonumber \\
f_R^i & \to & \bar u(p_4) \, \gamma_\mu \, (1+\gamma_5) \, u(p_2) \\
\bar f_L^i & \to & \bar v (p_2) \,  \gamma_\mu \, (1+\gamma_5) \, v (p_4) \, . \nonumber 
\end{eqnarray}
The cross sections for the  four  scattering processes (no average over helicities) are
\begin{equation}
\sigma\left(\nu_R f_L^i \to \nu_R f_L^i \right) =\frac{1}{3} \sigma\left(\nu_R \bar f_R^i \to \nu_R \bar f_R^i \right) = \frac{2}{3} \frac{G_i^2 s}{\pi} \quad ({\rm for \, bins} \, 
i= 
3,5)
\end{equation}
 and
\begin{equation}
\sigma\left(\nu_R \bar f_L^i \to \nu_R \bar f_L^i \right) = \frac{1}{3} \sigma(\nu_R f_R^i \to \nu_R f_R^i) = 
\frac{2}{3} \frac{G_i^2 s}{\pi} \quad ({\rm for \, bins}\, i = 1,2,4,6)  \, .
\end{equation}
In addition to these scattering processes, the $\nu_R$ interacts with the plasma through the annihilation processes: $\nu_R \bar\nu_L \to f_L^i \bar f_R^i,$ for  bins $i 
=3,5$, and
$\nu_R \bar\nu_L \to \, f_R^i \bar f_L^i$, for bins $i= 1,2,4,6.$ These all yield cross sections
$2 G_i^2 s/ (3\pi)$ due to  forward and backward suppression. Assuming all chemical potentials to be zero, the plasma will have an equal number density $n(T) = 
0.0913 
T^3$, for each fermion r.d.o.f. Thus,
\begin{equation}
\Gamma^{\rm scat} (T) = n(T) \left\langle \sum_{i=1}^6 \sigma_i(s) \, v_M \, {\cal N}_i \right \rangle \,, 
\end{equation}
where
$v_{\rm M} = 1- \cos \theta_{12}$ is the M\/{o}ller velocity, $s = 2 k_1 k_2 ( 1- \cos \theta_{12})$ is the square of the center-of-mass energy, and ${\cal N}_i$ is the 
multiplicty of Weyl fields in each bin (e.g., for $i =3, \,  {\cal N}_3 = 3 +2 = 5$). The scattering cross section is given by
\begin{equation}
\sigma_i^{\rm scat} = \sigma(\nu_R f_L^i \to \nu_R f_L^i) + \sigma(\nu_R \bar f_R^i \to \nu_R \bar f_R^i) = \frac{4}{3} \frac{2 G_i^2 \, s}{\pi} \quad {\rm for \, each} \, i = 
1, 
\dots, 6 \, ;
\label{15}
\end{equation}
similarly,
\begin{equation}
\sigma_i^{\rm ann} (s) = \sigma(\nu_R \bar \nu_L \to f_L^i \bar f_R^i + f_R^i \bar f_L^i) = \frac{1}{3} \frac{2 G_i^2 s}{\pi} \quad {\rm for \, each} \, i = 1, \dots 6 \, .
\label{16}
\end{equation}
Since $s = 2 k_1 k_2 (1 - \cos \theta_{12})$ and $v_M = 1 - \cos\theta_{12}$, we perform an approximate angular average $\langle (1- \cos\theta_{12})^2 \rangle = 4/3$, 
followed by a thermal averaging $\langle 2 k_1 k_2\rangle = 2 (3.15^2 \, T^2)$ to give
\begin{equation}
\Gamma^{\rm scat} (T)  = \left(\frac{4}{3}\right)^2 \, \frac{2}{\pi} \ 2  \ (3.15 T)^2 \, (0.0919 T^3)  \, \underbrace{\left(\sum_{i=1}^6 G_i^2 {\cal N}_i \right)}_{G_{\rm eff}^2} 
\simeq 2.05 G_{\rm eff}^2 \, T^5 \, .
\label{17}
\end{equation}
From (\ref{15}), (\ref{16}), and (\ref{17}),
\begin{equation}
 \Gamma^{\rm ann}(T)  = \frac{1}{4}  \Gamma^{\rm scat} (T)  \simeq 0.50 \ G_{\rm eff}^2 \ T^5 \, .
\end{equation}
Each of the $G_i$ is given by the sum of the contributions from $Z'$ and $Z''$ exchange,
\begin{equation}
4 \frac{G_i}{\sqrt{2}} = \frac{g'_6 \ g'_i}{M^2_{Z'}} + \frac{g''_6 \ g''_i}{M^2_{Z''}} \, .
\end{equation}

The Hubble expansion parameter during this time is
\begin{equation}
H(T) = 1.66 \ \langle N(T) \rangle^{1/2} \ T^2/M_{\rm Pl} \,,
\label{haim2}
\end{equation}
where $M_{\rm Pl}$ is the Planck mass. Since the quark-gluon energy density in the plasma has a similar $T$ dependence to that of the entropy (see Fig.~7 in~\cite{Bazavov:2009zn}), we take
$N(T) = 37~r(T)  + 19.5,$ so that $H(T) = 0.82\times 12.5~T^2/M_{\rm Pl}$. (The first factor provides an average for $r(T)$ over the temperature region, and we have now included the six $\nu_R$ r.d.o.f.) Since $\Gamma \propto T^5$ and $H \sim T^2$, it is 
clear that if at some temperature $T_{\rm dec}$, $H(T_{\rm dec}) = \Gamma_i (T_{\rm dec})$, the ratio $\Gamma/H$ will fall rapidly on further cooling. Thus from (\ref
{haim1}) and (\ref{haim2}) the equation determining $T_{\rm dec}$  depends on: {\em (1)} whether we need to preserve the absence of a chemical potential, or {\em (2)} 
whether we need simply to mantain physical equilibrium. The decoupling condition in these two cases is: {\em (1)} $\Gamma^{\rm ann} (T_{\rm dec}) = H (T_{\rm dec})$ 
and {\em (2)} $\Gamma^{\rm scat}  (T_{\rm dec})+ \Gamma^{\rm ann} (T_{\rm dec}) = H (T_{\rm dec})$; or numerically: {\em (1)}
\begin{equation}
0.50~G_{\rm eff}^2~T^5_{\rm dec} = 10.4~T_{\rm dec}^2/M_{\rm pl} \Rightarrow T^3_{\rm dec} = 20.8~(G^2_{\rm eff} M_{\rm Pl})^{-1} \,,
\end{equation}
and {\em (2)} 
\begin{equation}
2.50~G_{\rm eff}^2 \, T_{\rm dec}^5 = 10.4~T_{\rm dec}^2/M_{\rm Pl} \Rightarrow T_{\rm dec}^3 = 4.1~(G^2_{\rm eff} M_{\rm Pl})^{-1} \, .
\end{equation}
$T_{\rm dec}$ as determined from these equations must lie in the band
(\ref{10}).  

\begin{figure}[tbp]
\begin{minipage}[t]{0.49\textwidth}
\postscript{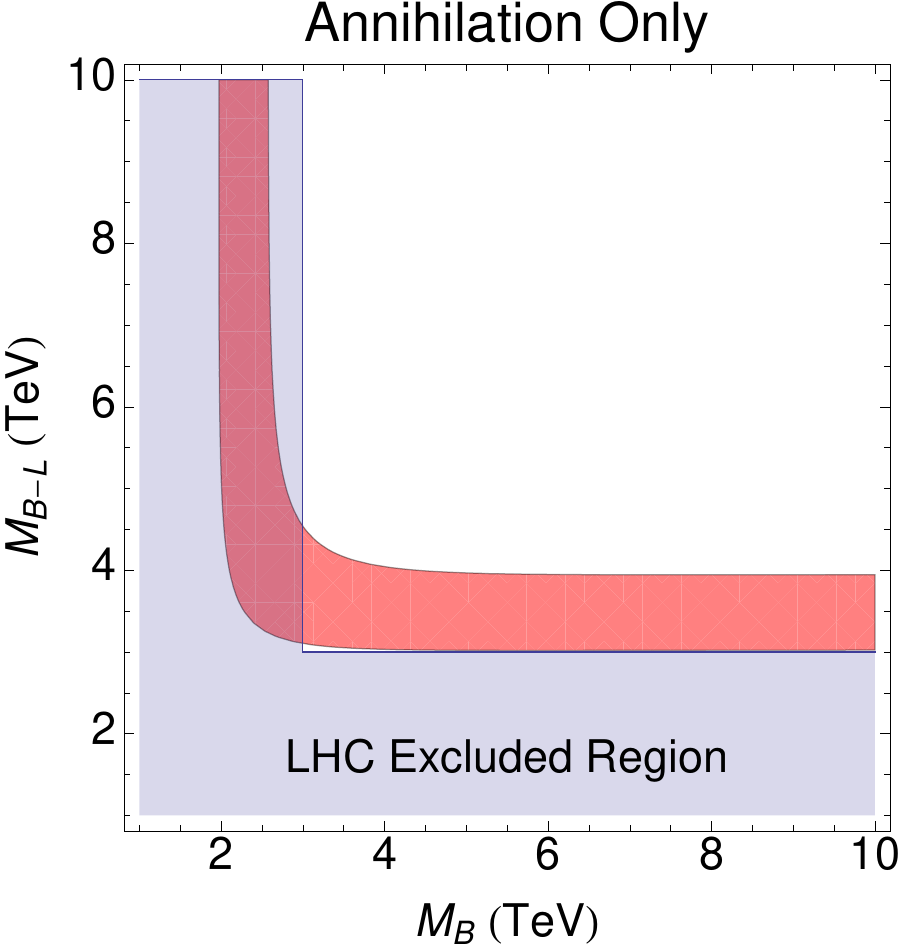}{0.99}
\end{minipage}
\hfill
\begin{minipage}[t]{0.49\textwidth}
\postscript{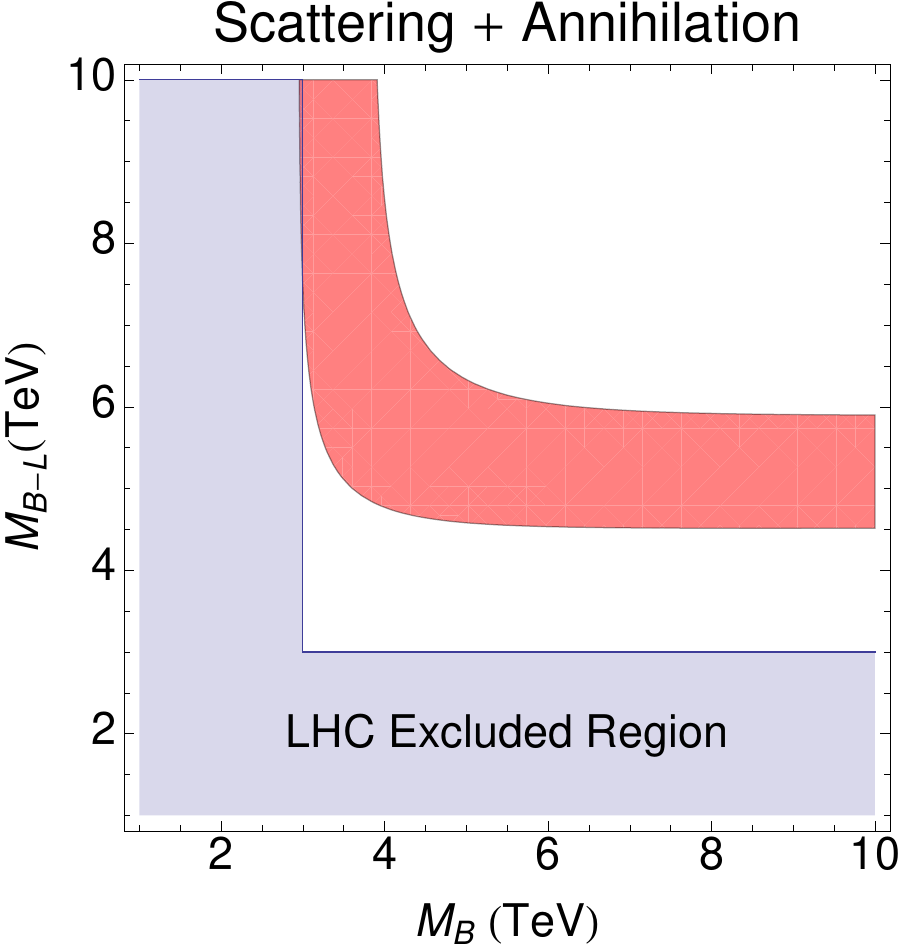}{0.99}
\end{minipage}
\caption{Dark shaded areas show region allowed from decoupling
  requirements to accommodate WMAP and BBN data. Light shaded regions
  indicate the masses excluded by the LHC7 dijet searches. The dark
  shaded areas in the left and right figures pertain to chemical and
  thermal equilibrium, respectively.  These two estimates should serve
  to bracket the size of the actual effect. The designation of $B$
  corresponds to $Z'$ and $B-L$ to $Z''$.}
\label{fig:rnus}
\end{figure}

Since all freedom of determining coupling constant and mixing angles has been exercised, there remains only constraints on the possible values of $M_{Z'}$ and $M_{Z''}$. Our results are encapsulated in Fig.~\ref{fig:rnus},  and along with other aspects of this work are summarized in these concluding remarks: 
\begin{itemize}
\item 
In this Letter, we develop a dynamic  explanation of recent hints that  the relativistic component of the energy during the era of last scattering is equivalent to about 1 extra Weyl neutrino.
\item We work within the context of a specific (string based) model with 3 $U(1)$ gauge symmetries, originally coupled to baryon number $B$,  lepton number $L$ , and a  3rd component of right-handed isospin $I_R$. We find that  rotation of the gauge fields to a basis exactly diagonal in hypercharge $Y$, and very nearly diagonal in $B-L$ and $B$ fixes all the mixing angles and the gauge couplings. Of course, of most significance for this work, requiring that the $B-L$ current be anomaly free, implies the  existence of 3 right-handed Weyl neutrinos.
\item We then find  that for certain ranges of $M_{B}$ and $M_{B-L}$ the decoupling of the $\nu_R$'s occurs during the course of the QCD phase transition, just so that they are only partially   reheated compared to the $\nu_L$'s  --- the desired outcome. 
\item To carry out this program, we needed to make use of some high statistics lattice simulations of a QCD plasma in the hot phase, especially the behavior of the entropy during the changeover.
\item Since our aim is to match the data, which has lower and upper bounds on the neutrino ``excess'', we obtain corresponding upper and lower bounds on the gauge field masses. Roughly speaking, if decoupling requires a freezout of the  annihilation channel (loss of chemical equilibrium), then 3 TeV $< M_{B-L}<$ 4 TeV. If thermal equilibrium via scattering is sufficient, then  4.5 TeV $< M_{B-L}<$ 6 TeV. These are ranges to be probed at LHC14.
\item Finally, a remark about the model: the gauging of $B$ allows a global conservation of baryon number. The gauging of $L$ brings with it the loss of a heavy Majorana for the seesaw model, as well as for leptogenesis through the decay of this particle. Thus, along with all its companion fields, the neutrino it is a Dirac particle, with the small mass originating through small Yukawa.
\end{itemize} 
 
L.A.A.\ is supported by the U.S.  National Science Foundation (NSF)  under  CAREER  Award PHY-1053663.  H.G.\ is supported by NSF Grant PHY-0757959.

\end{document}